\documentclass[twocolumn, tighten, times]{aastex631}

\usepackage{epstopdf}
\usepackage{amsmath}
\usepackage{enumitem}

\def\inn{_{\rm in}}
\def\out{_{\rm out}}

\submitjournal{ApJ}

\shorttitle{Secular planet-stirring of massive debris disks}
\shortauthors{A. A. Sefilian}

\begin{document}

\title{Massive Debris Disks May Hinder Secular Stirring by Planetary Companions: An Analytic Proof of Concept}

\author[0000-0003-4623-1165]{Antranik A. Sefilian}
\affil{Astrophysikalisches Institut und Universit{\"a}tssternwarte, Friedrich-Schiller-Universit\"at Jena, Schillerg\"a{\ss}chen~2--3, D-07745 Jena, Germany}
\affil{Center for Advanced Mathematical Sciences, American University of Beirut, PO Box 11-0236, Riad El-Solh, Beirut 11097 2020, Lebanon}
\affil{Alexander von Humboldt Postdoctoral Fellow. Email: \href{mailto:sefilian.antranik@gmail.com}{sefilian.antranik@gmail.com}}


\begin{abstract}
\noindent Debris disks or exo-Kuiper belts, detected through their thermal or scattered emission from their dusty components, are ubiquitous around main-sequence stars. Since dust grains are short-lived, their sustained presence is thought to require dynamical excitation, i.e., ``stirring", of a massive reservoir of large planetesimals, such that mutual collisions are violent enough to continually supply fresh dust. Several mechanisms have been proposed to explain debris disk stirring, with the commonly accepted being long-term, secular planet-debris disk interactions. However, while effective, existing planet-stirring models are rudimentary; namely, they ignore the (self-)gravity of the disk, treating it as a massless reservoir of planetesimals.  Here, using a simple analytical model, we investigate the secular interactions between eccentric planets and massive, external debris disks. We demonstrate that the disk gravity drives fast apsidal precession of both planetesimal and planetary orbits, which, depending on the system parameters, may well exceed the planet-induced precession rate of planetesimals. This results in strong suppression of planetesimal eccentricities and thus relative collisional velocities throughout the disk, often by more than an order of magnitude when compared to massless disk models. We thus show that massive debris disks may hinder secular stirring by eccentric planets orbiting near, e.g., the disk's inner edge, provided the disk is more massive than the planet. We provide simple analytic formulae to describe these effects. Finally, we show that these findings have important implications for planet inferences in debris-bearing systems, as well as for constraining the total masses of debris disks (as done for $\beta$ Pic).
\end{abstract}


\section{Introduction}
\label{sec:intro}

To date, around $20$ to $30$ per cent of main-sequence stars in the solar neighborhood have been found to harbor belts of debris, akin to Solar System's asteroid and Kuiper belts \citep[for recent reviews, see, e.g.,][]{hughes2018review, wyatt19review}. Such debris disks are detected through their excess thermal emission at infrared to millimeter wavelengths, and/or scattered light at optical to near-infrared wavelengths \citep[e.g.,][]{sibthorpe2018, Esposito2020}. These observations are typically explained by the presence of  sub-Earth mass populations of cm-sized and smaller dust grains \citep[e.g.,][]{wyattdent2002, holland17, krivovwyatt20}.

The dusty component of debris disks is considered to be of second generation \citep{wyatt08collisionsreview}: i.e., generated after the dispersal of protoplanetary disks, rather than being leftover from that phase. This is because dust grains are short-lived compared to the stellar age \citep{dd2003}, as various processes (such as radiation pressure and winds from the host stars) act to clear the dust population \citep{backman93}.  Thus, the sustained presence of dust grains in debris disks is attributed to a ``collisional cascade" involving large, km-sized planetesimals leftover from the protoplanetary disk phase \citep{wyatt08collisionsreview}. Within this picture, post-protoplanetary disk dissipation, collisions among large planetesimals generate smaller objects, perpetuating a cascade that continuously supplies fresh dust. To maintain the observed levels of dust, collisional models often require $\sim 10 - 100 M_{\earth}$ in parent planetesimals, and at times, surpassing $10^3 M_{\earth}$ -- a quantity that exceeds the total mass of solids thought to be present during the protoplanetary disk phase. This issue is recognized as the ``debris disk mass problem" \citep[][]{krivovwyatt20}.

For a collisional cascade to be ignited, the reservoir of parent planetesimals needs to satisfy two conditions. First, that parent planetesimals are abundant enough to ensure collisions are frequent, and second, that their orbits are dynamically excited with high enough relative velocities to ensure collisions are destructive \citep{dohnanyi1969}. While planetesimal formation is an active area of research \citep[e.g.,][]{Birnstiel}, it is usually assumed that planetesimals form on cold, near-coplanar and circular orbits \citep[e.g.,][]{Kenyon2008}. Accordingly, it is often thought that additional means of dynamical excitation of planetesimals is required for debris disks to become visible. This process is referred to as ``stirring'' in the literature \citep{wyatt08collisionsreview}.

Two key mechanisms for stirring have been proposed in the literature, apart from the possibility that debris disks are already \textit{pre-stirred} by processes acting during the protoplanetary disk phase \citep{wyatt08collisionsreview, Matthews2014, Najita2022}. One is known as \textit{self-stirring} \citep[][]{Kenyon2008, Kennedy2010}, whereby planetesimals ignite a collisional cascade due to their gravitational perturbations among each other. This process requires the growth of some planetesimals to a size comparable to Pluto's, which then dynamically excite the remaining smaller ones. A second mechanism is known as \textit{planet-stirring}, whereby planets excite planetesimal orbits through gravitational interactions, igniting a collisional cascade. Various flavors of this mechanism have been explored, including those examining the long-term, secular perturbations caused by eccentric planets \citep{Mustill2009}, and those focused on short-term perturbations due to circular planets, such as mean-motion resonances and scattering \citep{Tyson2023}. In a recent study, \citet{Munoz2023} investigated the synergy of planet- and self-stirring, termed \textit{mixed stirring}, finding enhanced efficiency compared to isolated cases of either mechanism.

Each stirring mechanism comes with its own set of advantages and drawbacks \citep{wyatt08collisionsreview}. For instance, while self-stirring does not invoke any planets, it may necessitate considerably large disk masses for optimal efficiency. Indeed, recently, \citet{Pearce2022JWST} have shown that for the 67 resolved debris disks they consider, the median minimum mass required for self-stirring is $110 M_{\earth}$, among which $17$ disks require masses in excess of $10^3 M_{\earth}$ (within a 1-sigma confidence level; see their Table 1). This is deemed to be unfeasible, as such masses exceed the total solid mass thought to be present during the protoplanetary disk phase \citep[$\approx 10^{3} M_{\earth}$;][]{krivovwyatt20}. Accordingly, some recent studies tend to favor planet-stirring models, in one or another flavor  \citep[e.g.,][]{Pearce2022JWST, Munoz2023, Tyson2023}.
We note, however, that this conclusion is open to question; after all,  mass estimates are subject to uncertainties, such as those related to the initial size distribution of debris particles. Thus, the masses required  for self-stirring  may be different, and possibly lower, than commonly thought for different,  perhaps more plausible, size distributions \citep[see, for instance,][]{Najita2022}.

In this work, we focus on planet-stirring models. An important aspect absent in planet-stirring models is the dynamical role of the disk's gravity, which, in principle may affect the evolution of both the planetesimals and the planet. That is, planet-stirring models often treat the debris disk as a collection of massless planetesimals, solely influenced by the gravitational forces exerted by the star and the (invoked) planets. This raises several questions, motivating the present work: How does the disk gravity affect existing models of planet-stirring, particularly in the context of secular stirring? Can debris particles ``resist'' planet-stirring through their collective gravity? If so, how does this impact planet inferences derived from simple, massless disk models?

In the realm of protoplanetary disks, studies have already explored somewhat similar questions. For instance, \citet{rafikov_ptype, rafikov_stype} was first to show that massive, \textit{axisymmetric} protoplanetary disks can counteract perturbations due to eccentric stellar companions in S- and P-type systems. This could help in, e.g., overcoming the infamous fragmentation barrier by preventing planetesimals from being compelled into highly eccentric and destructive orbits  \citep[see also,][]{sil15, silsbeekepler,silsbee2021, sefilian-msc, marcy2024}, provided that the protoplanetary disk remains nearly axisymmetric \citep[see, e.g.,][]{marzari2009, marzari2012}.
A related study by \citet{bat11} has shown that massive protoplanetary disks can suppress, and in some cases, inhibit the Kozai-Lidov oscillations expected from a distant stellar companion on an inclined orbit. The common denominator in these studies is the \textit{disk-driven} fast apsidal precession of the planetesimal and/or the stellar companion orbits.

In the realm of debris disks, however, disk (self-)gravity is often not considered \citep[e.g.,][to name a few]{wyattetal99, leechiang2016, nesvold17, Pearce2022JWST, farhat-sefilian-23}. 
Indeed, with the exception of a few studies that do account for the back-reaction of the disk on the planet \citep[e.g.,][]{beust2014, pearce2015}, most ignore the  gravitational effects of the disk on itself \citep[see, however,][]{Pedro2023}.
Recently, using (semi-)analytical methods, \citet{Paper1, Paper2} presented the first detailed investigation of long-term, secular interactions between eccentric planets and external, \textit{self-gravitating} debris disks. They examined scenarios involving disks that are less massive than the planets, finding that a robust outcome is the formation of a wide gap within the disk, giving rise to `double-belt' structures akin to those observed in several systems \citep[see, e.g.,][]{marino2022review}.  In \citet{Paper1}, the authors also noted the potential of relatively massive debris disks to suppress eccentricity excitation due to planets, albeit only in passing.

In this work, motivated by the pioneering works of \citet{rafikov_ptype, rafikov_stype}, we propel from the observations made by \citet{Paper1} and show that in single-planet systems with a more massive debris disk than the planet, the disk gravity can reduce planetesimal eccentricities across the disk. Accordingly, the relative velocities at which planetesimals collide are reduced, often by more than an order of magnitude when compared to massless debris disk models. This implies that  debris disks, if massive enough, may hinder secular stirring by planetary companions on eccentric orbits. We now describe this idea and its implications in more detail.

\section{Model System}
\label{sec:modelsystem}

Our general setup is similar to that explored in \citet{Paper1}. We consider a coplanar system comprising of a central star of mass $M_c$, a single planet of mass $m_p$, and a debris disk of mass $M_d$  situated exterior to the planet (with $m_p, M_d \ll M_c$). The planet's orbit is assumed to be initially eccentric, characterized by its semimajor axis $a_p$, eccentricity $e_p$ (typically, $e_p \lesssim 0.1$), and apsidal angle $\varpi_p$. The debris disk is assumed to be razor-thin,  and initially circular and axisymmetric. Its inner edge $a\inn$ is set to be at $ 5$  Hill radii $r_{\textrm{H,Q}}$ from the planet's apocenter, $Q_p = a_p (1+e_p)$, where \citep[][]{pearcewyatt2014}:
\begin{eqnarray}
    r_{\textrm{H, Q}} 
    &\approx&  Q_p
    \bigg[ \frac{m_p}{(3-e_p) M_c} \bigg]^{1/3} ,
    \label{eq:rHQ}
\end{eqnarray}
and the outer edge is set at $a\out = 5 a\inn$. The choice of $a\inn =Q_p + 5 r_{\textrm{H, Q}}$ is motivated by the fact that an eccentric planet is expected to clear planetesimals within $5$ Hill radii of its apocenter \citep[e.g.,][]{pearce2024}, while $\delta \equiv a\out/a\inn =5$ is chosen to ensure a radially-broad enough disk.

We further characterize the disk's surface density $\Sigma_d(a)$ by a truncated power-law profile with an exponent $p$, such that
\begin{eqnarray}
    \Sigma_d(a) &=& \Sigma_0 \bigg( \frac{a\out}{a} \bigg)^p 
    \label{eq:Sigma_d}
\end{eqnarray}
for $a\inn \leq a \leq a\out$, and $\Sigma_d(a) = 0 $ elsewhere. Accordingly, the total mass $M_d$ of the debris disk, given by  $M_d = 2\pi \int_{a\inn}^{a\out} a \Sigma_d(a) da$, reads as:
\begin{eqnarray}
    M_d &=& \frac{2\pi}{2-p} \Sigma_0   a\out^2 
    \big( 1- \delta^{p-2}   \big) . 
    \label{eq:Mdisk}
\end{eqnarray}
From hereon, we adopt a fiducial disk model with $p=3/2$, which corresponds to the slope of the Minimum Mass Solar Nebula \citep{hayashi}.

\section{Basic Physics: Secular Perturbations}
\label{sec:basicphysics}

We are interested in the long-term dynamics of large, $\sim$km-sized planetesimals initiated on circular orbits within the debris disk. We characterize a planetesimal's orbit with its semimajor axis $a$, mean motion $n = \sqrt{GM_c/a^3}$, eccentricity $e$, and  apsidal angle $\varpi$. Given the insensitivity of large planetesimals to radiative non-gravitational forces \citep{burns79}, we focus  purely on perturbative effects arising due to the gravity of \textit{both} the disk and the planet.

To this end, we follow the calculations presented in \citet{Paper1}, performed within the framework of orbit-averaged, secular perturbation theory. Considering the model system of Section \ref{sec:modelsystem}, \citet{Paper1} presented an analytical expression for the secular disturbing function $R$ -- valid to second order in eccentricities\footnote{For a non-linear treatment, valid to fourth order in eccentricities, we refer the reader to, e.g., the study of \citet{ST19} concerning the secular interactions between a massive, self-gravitating trans-Neptunian disk and the giant planets in the Solar System \citep[see also,][]{wardhahn1998}.} -- accounting for (i) the disk's gravity, acting \textit{both} on the planet and the planetesimals,  and (ii) the planet's gravity acting on the planetesimals; see their Equation (9). These calculations are done by considering the axisymmetric component of the disk gravity, but ignoring its non-axisymmetric contribution. To streamline our conceptual framework presentation, we adhere to this assumption in our present work, reserving the discussion of potential consequences upon its relaxation for Section \ref{sec:discussion}.

\citet{Paper1} then derived the equations of motion describing the evolution of the planetesimal eccentricity vector, namely, $\mathbf{e} = (K, H) = e (\cos\Delta\varpi, \sin\Delta\varpi)$ with $\Delta \varpi \equiv \varpi - \varpi_p$,  finding (see also their Equation (10)):
\begin{eqnarray}
    \frac{dK}{dt} &\approx&  \frac{-1}{n a^2}\frac{\partial R}{\partial H} = - (A - A_{d,p}) H,  
    \nonumber
    \\
    \frac{dH}{dt} &\approx&   \frac{1}{n a^2} \frac{\partial R}{\partial K} = (A - A_{d,p}) K + B_p     .
    \label{eq:EOM}
\end{eqnarray}
These equations of motion are derived in a frame co-precessing with the planetary orbit, thus the definition of $\mathbf{e}$ in terms of $\Delta\varpi$ (rather than $\varpi$). This is done because the disk gravity causes the planet's apsidal angle to precess linearly in time at a constant rate of $A_{d,p}>0$, i.e., $\varpi_p(t) = A_{d,p} t + \varpi_p(0)$ (from hereon, $\varpi_p(0)=0$).

The expressions for the various coefficients appearing in Equation (\ref{eq:EOM}) are given  in \citet[][see their Equations (4)--(8)]{Paper1}, and we outline them below for completeness. The term $A_{d,p}$ is given by \citep[see also,][]{petrovich19}: 
\begin{eqnarray}
    A_{d,p}  &=& \frac{3}{4} n_p \frac{2-p}{p+1} \frac{M_d}{M_c} \bigg(\frac{a_p}{a\out}\bigg)^3 \frac{\delta^{p+1}-1}{1-\delta^{p-2}}\phi_1^c  , 
    \label{eq:Adp} 
    \\
    &\approx&  7.04 \times 10^{-1}  ~ \textrm{Myr}^{-1} 
        ~ \frac{\phi_1^c}{3}
    ~ \bigg( \frac{M_d}{1  ~ M_N} \bigg)
    ~ \frac{a_{p,30}^{3/2} }{a_{\textrm{out}, 150}^{3}}
    ~  M_{c,1}^{-1/2} . \nonumber
\end{eqnarray}
In Equation (\ref{eq:Adp}), $n_p = \sqrt{G M_c / a_p^3}$ is the planetary mean motion,  $M_N$ is Neptune mass, and we have defined: $a_{p, 30} \equiv a_p/(30 {\textrm{au}})$, $a_{\textrm{out}, 150} \equiv a\out/(150 {\textrm{au}})$, and $M_{c, 1} \equiv M_c/(1 M_{\odot})$. The numerical estimate in Equation (\ref{eq:Adp}) is for $p =3/2$ and $\delta =5$, assuming $a_p/a\inn \approx 0.8$ so that $\phi_1^c \approx 3$. Here, $\phi_1^c$ is a dimensionless factor of order unity given by Equation (A7) in \citet{Paper1}; it is a function of  $a_p/a\inn$ that varies very weakly with $p$ and  $\delta$ (see also their Figure 13). Indeed, irrespective of $(p, \delta)$,  one has $\phi_1^c \approx 1$ for $a_p/a\inn \ll 1$, and  $\phi_1^c \sim 10$ for $a_p/a\inn \sim 1$.

In Equation (\ref{eq:EOM}), the term $A \equiv A_p + A_d$ represents the free precession rate of the planetesimal, contributed both by the gravity of the planet $A_p$ and the disk $A_d$. 
The expression for $A_p$ is given by \citep[see also,][]{mur99}:
\begin{eqnarray}
    A_p &=& \frac{1}{4}  n \frac{m_p}{M_c} \frac{a_p}{a} b_{3/2}^{(1)}(a_p/a),      \label{eq:Ap} 
    \\
    &\approx&  1.30 \times 10^{-1}  ~ \textrm{Myr}^{-1}  
    ~ \bigg( \frac{m_p}{1~ M_N} \bigg)
    ~ \frac{ a_{p, 30}^2 } { a_{60}^{7/2}}
    ~ M_{c,1}^{-1/2}      ,  
    \nonumber 
\end{eqnarray}
where we have defined $a_{60} \equiv a/(60 {\textrm{au}})$, and $A_d$ reads as:
\begin{eqnarray}
    A_d &=&        (2-p) n \frac{M_d}{M_c} \bigg( \frac{a}{a\out} \bigg)^{2-p} \frac{\psi_1}{1-\delta^{p-2}} , 
    \label{eq:Ad} 
    \\
    &\approx& 
    -2.19 \times 10^{-1}
    ~ \textrm{Myr}^{-1}  
    ~ \bigg| \frac{\psi_1}{-0.55}  \bigg|
        ~ \bigg( \frac{M_d}{1~ M_N} \bigg)
    ~ \frac{a_{60}^{-1}}{a_{\textrm{out}, 150}^{1/2}}
         ~  M_{c,1}^{-1/2}   
    , \nonumber 
\end{eqnarray}
see also \citet{sil15}. In Equation (\ref{eq:Ap}), $b_s^{(m)}(\alpha)$ is the standard Laplace coefficient \citep{mur99},
\begin{equation}
    b_s^{(m)}(\alpha) = \frac{2}{\pi} \int_{0}^{\pi} \cos(m\theta) \bigg( 1+\alpha^2-2\alpha \cos\theta \bigg)^{-s}   d\theta, 
\end{equation}
and the numerical estimate assumes $b_{3/2}^{(1)}(\alpha) \approx 3 \alpha$, valid for $\alpha \ll 1$. The term $\psi_1$ appearing in Equation (\ref{eq:Ad}) is a dimensionless coefficient of order unity that depends on the disk's surface density slope $p$ and the planetesimal's semimajor axis relative to the disk edges \citep{sil15, SR19}. For a $p=3/2$ disk, $\psi_1(a) \approx -0.55$ for $a\inn \ll a \ll a\out$, and $\psi_1(a)$ diverges as $a \rightarrow a\inn, a\out$. Note that the disk drives retrograde planetesimal precession, $A_d(a) < 0$,  opposite to the planet's effect, $A_p(a)>0$. Moreover, it is important to note that for the particular set of parameters in Equations (\ref{eq:Adp})--(\ref{eq:Ad}) with $m_p = M_d$ and $a_p/a\inn \approx 0.8$, one has $A_{d,p} \gtrsim |A_{d}| \gtrsim A_p$.

Finally, the term $B_p$ in Equation (\ref{eq:EOM}) represents the excitation of planetesimal eccentricity driven by the eccentric planet's non-axisymmetric potential. It is given by:
\begin{eqnarray}
        B_p &=& -\frac{1}{4} n \frac{m_p}{M_c} \frac{a_p}{a} b_{3/2}^{(2)}(a_p/a) e_p .
        \label{eq:Bp}
\end{eqnarray}
Note that the analogous term due to the disk -- which may arise once the disk develops some degree of non-axisymmetry -- is ignored.

This completes our description of the analytical framework. Before moving on, we remark that this framework is similar to that first developed by \citet{rafikov_ptype, rafikov_stype} for studying the dynamics of planetesimals and its implications for planet formation in stellar binaries \citep[see also,][]{sil15, silsbeekepler, silsbee2021}.

\section{Planetesimal  Dynamics in massive disks}
\label{sec:dynamics}

We now analyze the secular dynamics of planetesimals in the combined planet--disk potential. Assuming initially circular orbits for the planetesimals, i.e., $K(0) = H(0) = 0$,  Equation (\ref{eq:EOM}) admits an analytical solution given by:
\begin{eqnarray}
    e(t) &=& e_m \bigg| \sin\bigg( \frac{A_p + A_d -A_{d,p}}{2} t \bigg)   \bigg|
    , 
    \label{eq:e_t}
    \\
    \tan [\Delta\varpi(t)] &=& \tan\bigg( \frac{A_p + A_d -A_{d,p}}{2} t - \frac{\pi}{2}  
  \bigg), 
  \label{eq:w_t}
\end{eqnarray}
where $\Delta\varpi$ stays in the range $[-\pi, \pi]$; see also Equations (11) and (12) in \citet{Paper1}. Here, $e_m(a) = 2|e_{\textrm{f}}(a)|$ is the maximum amplitude of eccentricity oscillations characterized by a period of $T_{\rm{osc}} = 2\pi/|A_p + A_d -A_{d,p}|$, where $e_{\textrm{f}}(a)$ is the forced eccentricity:
\begin{eqnarray}
    e_{\textrm{f}}(a) = \frac{-B_p(a)}{A_p(a) + A_d(a) - A_{d,p}} . 
    \label{eq:e_forced}
\end{eqnarray}
Note that in  a massless disk one has $A_d = A_{d,p}  = 0$, and Equations (\ref{eq:e_t})--(\ref{eq:e_forced}) reduce to the classical results for test particles perturbed by an eccentric, non-precessing planet \citep{mur99}. Equations (\ref{eq:e_t})--(\ref{eq:e_forced}) have been previously utilized by \citet{Paper1} to study planetesimal dynamics in debris disks with $M_d/m_p \lesssim 1$. In the present work, we focus on the case where $M_d/m_p \gtrsim 1$.

\subsection{Precession rates}
\label{sec:precession_rates}

Equations (\ref{eq:e_t})--(\ref{eq:e_forced}) show that the secular evolution of planetesimals depends on their  precession rate $A(a)$ \textit{relative} to the planet $A_{d,p}$. To assess the disk's contribution, we measure the disk-driven precession rates  of planetesimals $A_d$ and the planet $A_{d,p}$ relative to the planet-driven planetesimal precession $A_p$. Using Equations  (\ref{eq:Adp})--(\ref{eq:Ad}), we find that:
\begin{eqnarray}
    \frac{|A_d|}{A_p} 
    &\approx& \frac{4 |(2-p) \psi_1|}{3(\delta^{2-p}-1)} \frac{M_d}{m_p} \bigg(\frac{a}{a\inn}\bigg)^{4-p} \bigg(\frac{a\inn}{a_p}  \bigg)^{2} , 
    \label{eq:Ad_Ap}
    \\
      &\approx&  1.28
    ~ \bigg|\frac{\psi_1}{-0.55}\bigg| 
    ~ \frac{M_d}{m_p} 
    ~ \bigg( \frac{a/a\inn}{1.5}\bigg)^{5/2} 
    ~ \bigg( \frac{a\inn/a_p}{1.25}\bigg)^{2} , 
    \nonumber
\end{eqnarray}
and
\begin{eqnarray}
    \frac{A_{d,p}}{A_p} 
    &\approx& \frac{(2-p) \phi_1^c}{(p+1) C \delta^3} \frac{M_d}{m_p} \bigg(\frac{a}{a\inn}\bigg)^{7/2}  \bigg( \frac{a\inn}{a_p}\bigg)^{1/2}
    , 
    \label{eq:Adp_over_Ap}
    \\
   &\approx& 2.20
     ~ \frac{\phi_1^c  }{3}
     ~ \frac{M_d}{m_p} 
     ~ \bigg(\frac{a/a\inn}{1.5}\bigg)^{7/2}
     ~ \bigg( \frac{a\inn/a_p}{1.25}\bigg)^{1/2} . 
    \nonumber
\end{eqnarray}
Here, the numerical estimates assume $p = 3/2$ and $\delta = 5$ and, for brevity, we have defined $C \equiv (1-\delta^{p-2})/(\delta^{p+1}-1)$.

\begin{figure}[t!]
\epsscale{1.17}
\plotone{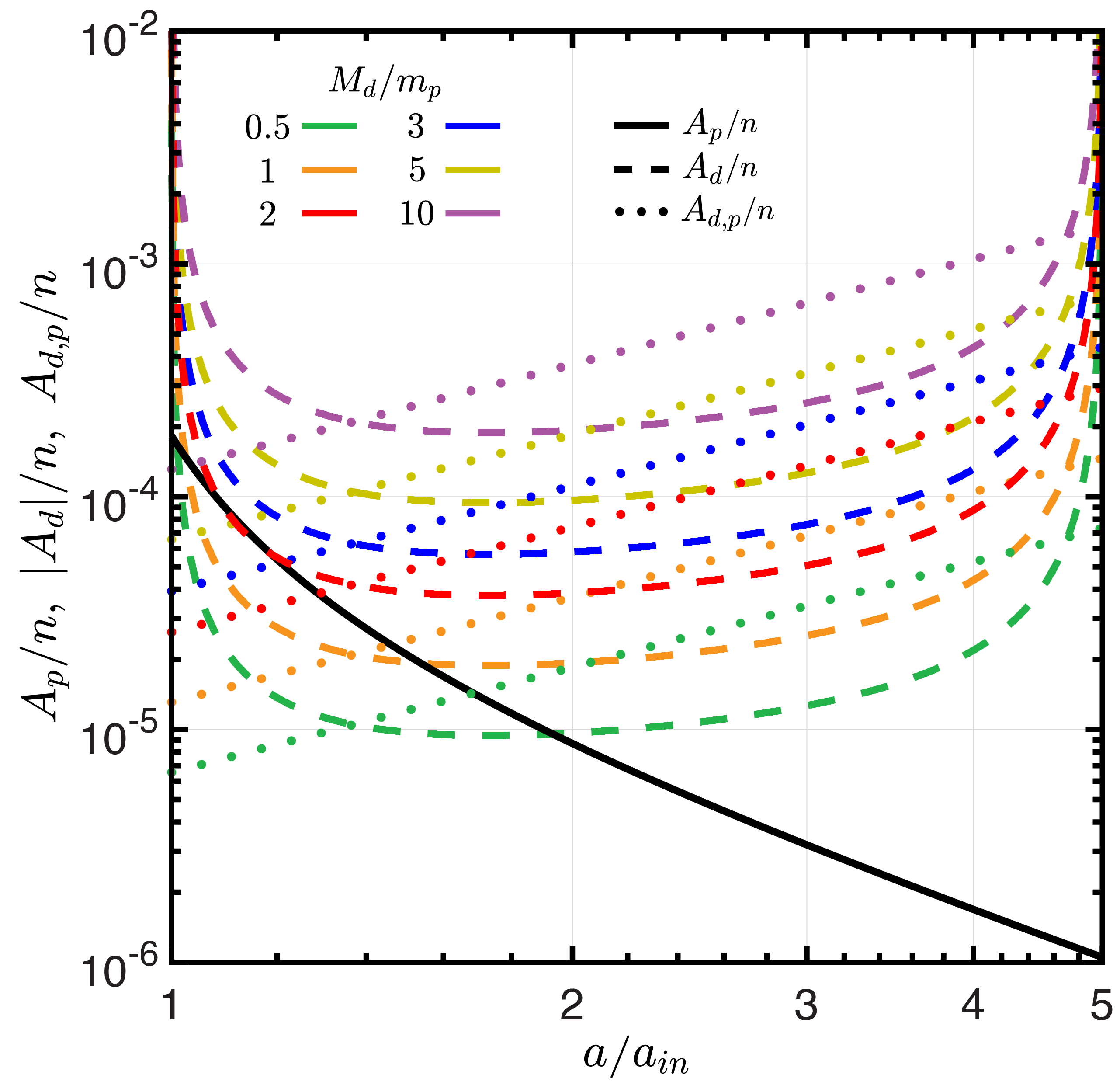}
\caption{Profiles of the characteristic precession rates in the problem. The plot shows $A_p/n$ (Equation (\ref{eq:Ap}); solid black line), $|A_d|/n$ (Equation (\ref{eq:Ad}); dashed lines), and $A_{d,p}/n$ (Equation (\ref{eq:Adp}); dotted lines), as a function of planetesimal semimajor axis $a/a\inn$. Calculations for $A_d$ and $A_{d,p}$ assume different values of $M_d/m_p$ (shown with different colors), assuming $M_c = 1M_{\odot}$ and a Neptune-mass planet at $a_p = 30$ au with $e_p=0.1$. The latter sets the disk's inner at $a\inn \approx 1.25 a_p$ (Equation \ref{eq:rHQ}), assumed to have $p=3/2$ (Equation \ref{eq:Sigma_d}). It is evident that for $M_d/m_p > 1$, the disk-driven precession rates, i.e., both $A_d$ and $A_{d,p}$, dominate throughout the entire disk: namely, $|A_d| \gtrsim A_p, A_{d,p}$ within $a/a\inn \sim 1.5$, and $A_{d,p} \gtrsim |A_d|, A_p$ beyond that region; in line with Equations (\ref{eq:critical_a_for_Ad_Adp}) and (\ref{eq:Mdisk_condition}). See the text (Section \ref{sec:precession_rates}) for details.}
\label{fig:precession-rates}
\end{figure}

Equations (\ref{eq:Ad_Ap}) and (\ref{eq:Adp_over_Ap}) show that both ratios are increasing functions of $a/a\inn$. This holds true for all astrophysically motivated values of $p$, i.e., $ 0 < p < 2$, so that more mass is concentrated in the outer parts of the disk than in the inner regions (see Equations (\ref{eq:Sigma_d}) and (\ref{eq:Mdisk})). Thus, for the fiducial set of parameters adopted here (Section \ref{sec:modelsystem}), the disk's (self-)gravitational effects would dominate over that due to the planet at distances beyond $a/a\inn \sim 1.5$,  such that $ A_{d,p}/A_p \gtrsim |A_d|/A_p \gtrsim 1$. More specifically, we find that  $A_{d,p} \gtrsim |A_d|$ for $a \gtrsim a_c$ where
\begin{eqnarray}
   a_c &\approx&   1.43  a\inn  ~ \bigg( \frac{a\inn/a_p}{1.25} \bigg)^{3/2}
   ~ \bigg| \frac{\psi_1}{-0.90} \bigg|
   ~ \frac{3}{\phi_1^c},
    \label{eq:critical_a_for_Ad_Adp}
\end{eqnarray}
i.e., practically for any $a \gtrsim a\inn$ as long as $a_p \sim a\inn$.\footnote{Note that, in practice, $|\psi_1(a)|$ varies between $\approx 0.95$ and $\approx 0.83$ within $a/a\inn = 1.3$ and $1.4$, which is where $A_{d,p} = |A_d|$ according to Figure \ref{fig:precession-rates}.} 
Note that $a_c$ is not dependent on the involved masses  but is instead strongly influenced by $a_p$.
Within  $a  \lesssim a_c$, on the other hand, $A_p$ would be the dominant secular frequency, except if $|A_d| > A_p$ which, according to Equation (\ref{eq:Ad_Ap}) evaluated at $a = a\inn$, requires
\begin{eqnarray}
    \frac{M_d}{m_p} &\gtrsim& 1.32   ~ \bigg| \frac{-0.90}{\psi_1} \bigg| ~ \bigg( \frac{a\inn/a_p}{1.25} \bigg)^{-2}  . 
    \label{eq:Mdisk_condition}
\end{eqnarray}
Disk-to-planet mass ratios smaller than that in Equation (\ref{eq:Mdisk_condition}) would lead to the occurrence of secular resonances near $a\approx a\inn$ where $A(a) \approx A_p = A_{d,p}$ \citep{Paper1}, and $A_{d,p}$ would dominate the dynamics beyond that location.

To illustrate this further,  in Figure \ref{fig:precession-rates} we plot the behavior of $A_p$, $|A_d|$, and $A_{d,p}$ (scaled by $n$) as a function of planetesimal semimajor axis for several values of $M_d/m_p$. Calculations assume a Neptune-mass planet at $a_p = 30$ au with $e_p = 0.1$ around a Solar-mass star such that, according to Equation (\ref{eq:rHQ}), the disk's inner edge is at  $a\inn\approx 1.25 a_p \approx 37$ au (with $\delta = 5$ and $p=3/2$; Section \ref{sec:modelsystem}). Figure \ref{fig:precession-rates} shows that for $0.5 \lesssim M_d/m_p \lesssim  10$,  the disk-driven \textit{planetary} precession is indeed the dominant secular frequency beyond $a/a\inn \sim 1.5$, such that $A_{d,p} \gtrsim |A_d|\gtrsim A_p$ (in line with Equation (\ref{eq:critical_a_for_Ad_Adp})). This implies the counter-intuitive result that the disk's gravity acting on the planet is more important than its direct effect on the planetesimals embedded within, i.e., $A_{d,p} \gtrsim |A_d|$. An exception to this can be seen at $a\sim a\inn, a\out$ (regardless of $M_d/m_p$), where $|A_d|$ dominates. This, however, is due to the divergence of $A_d \propto -|\psi_1|$ for $a\rightarrow a\inn, a\out$ (Eq. (\ref{eq:Ad}); see also \citet{sil15, SR19}).

Interior to  $ a/a\inn \lesssim 1.5$, Figure \ref{fig:precession-rates} shows that the planet-driven planetesimal precession rate dominates over other frequencies such that $A_{p} \gtrsim A_{d,p},|A_d|$, i.e., as in the massless disk case, only when $M_d/m_p \lesssim 1$ (see also Eq. (\ref{eq:Mdisk_condition})). The transition from planet- to disk-dominated regimes occurs via a secular resonance where $A_p + A_d = A_{d,p}$ and  $e_{\textrm{f}}$ diverges (see Eq. (\ref{eq:e_forced}), Figure \ref{fig:eccentricities} and \citet{Paper1}). For much more massive disks with  $M_d/m_p \gtrsim 1$, on the other hand, Figure \ref{fig:precession-rates} shows that the disk-driven planetesimal precession rate $A_d$ (instead of $A_p$) would be dominant in the innermost regions, $a \lesssim 1.5 a\inn$, in line with  Equation (\ref{eq:Mdisk_condition}). This implies that when $M_d/m_p > 1$, the \textit{relative} precession rate of planetesimal orbits  would be dominated by the disk (self-)gravity alone at all semimajor axes $a\inn \leq a \leq a\out$ (i.e., $A(a) - A_{d,p} \approx A_d(a) - A_{d,p}$), while the planet acts only with its non-zero non-axisymmetric torque (i.e., $B_p$, Equation (\ref{eq:Bp})). It is this regime that we further explore in this study.

Note that the behavior described above holds as long as $a_p \sim a\inn$, as would be the case if, e.g., the planet is responsible for sculpting the disk's inner edge (Equation \ref{eq:rHQ}). Otherwise, if $a_p \ll a\inn$ (not shown in Figure \ref{fig:precession-rates}), one has $A_{d,p} \propto a_p^{3/2} \rightarrow 0$ (Equation (\ref{eq:Adp})), and the disk-driven planetesimal precession rate $A_d(a)$ (rather than $A_{d,p}$) would dominate beyond some critical semimajor axis. Generally, for a given planet with  $a_p/a\inn \lesssim 1$, this may happen for all values of $M_d/m_p \lesssim 1$. The details of this case are omitted here; for an in-depth exploration, we refer the reader to  \citet{Paper1, Paper2} and \citet{sefilian-phd}.

\subsection{Planetesimal eccentricities}
\label{sec:planetesimal_eccentriticies}

Section \ref{sec:precession_rates} suggests that when considering secular planet-stirring further away from the planet, i.e., typically at  $a\gtrsim a_c$ (Equation (\ref{eq:critical_a_for_Ad_Adp})), one can neglect the planetesimal precession rates due to the planet $A_p$ and the disk self-gravity $A_d$ compared to the disk-driven \textit{planetary} precession $A_{d,p}$ provided that $M_d/m_p \gtrsim 1$ (see also Eq. (\ref{eq:Mdisk_condition}) and Figure \ref{fig:precession-rates}). Accordingly, Equation (\ref{eq:e_forced}) predicts that when $A_{d,p} \gg A_p,|A_{d}|$, the forced planetesimal eccentricities  $e_{\rm f} \rightarrow e_{\rm f}^{d}$ so that:
\begin{eqnarray}
    e_{\rm f}^{d} 
    =  \bigg| \frac{B_p}{A_{d,p}}\bigg| 
    &\approx&  \frac{5 (p+1)  C  }{4 |(2-p)| \phi_1^c}  e_p  
    \frac{m_p}{M_d}
    \bigg( \frac{a_p}{a\out} \bigg)^{3/2}
    \bigg( \frac{a\out}{a}  \bigg)^{9/2}, 
    \label{eq:ef_diskplanet}
    \\
    &\approx& 1.16 \times 10^{-3} ~ 
    \frac{3}{\phi_1^c}
    ~ \frac{m_p}{M_d} ~ \frac{e_p}{0.10} ~ \frac{a_{p,30}^{3/2} ~ a_{{\textrm{out}, 150}}^3}{a_{100}^{9/2}}
    . \nonumber 
\end{eqnarray}
Here, as before, $C = (1-\delta^{p-2})/(\delta^{p+1}-1)$, we have approximated $b_{3/2}^{(2)}(\alpha) \approx (15/4) \alpha^2$ (valid for $\alpha \ll 1$), and the numerical estimate assumes $p = 3/2$ and $\delta = 5$. Equation (\ref{eq:ef_diskplanet}) is similar to the one derived by \citet{rafikov_ptype} for planetesimals in massive, circumbinary protoplanetary disks.\footnote{An expression that is somewhat similar to Equation (\ref{eq:ef_diskplanet}) is also obtained by \citet[][]{Paper1}, however, assuming $|A_d| \gg A_{d,p}, A_p$ (rather than $A_{d,p} \gg |A_{d}|, A_p$ as done here). In that case, which is  valid for $a_p/a\inn \ll 1$, one instead finds $e_{\textrm{f}}^d \propto a_p^{3}/a^{5-p}$; see their Equation (15).}

Equation (\ref{eq:ef_diskplanet}) is to be compared with the forced planetesimal eccentricities in a massless disk, in which case $e_{\rm f} \rightarrow e_{\rm f}^{p}$:
\begin{eqnarray}
    e_{\rm f}^{p} = \bigg| \frac{B_p}{A_{p}}\bigg| 
    = \frac{b_{3/2}^{(2)}(a_p/a)}{b_{3/2}^{(1)}(a_p/a)} e_p 
   & \approx & \frac{5}{4} \frac{a_p}{a} e_p  , 
    \label{eq:ef_planet}
    \\
  &  \approx &
    3.75 \times 10^{-2}  ~\frac{e_p}{0.10}  ~ \frac{a_{p,30}}{a_{100}},
\nonumber
\end{eqnarray}
see also \citet{Mustill2009}. Equations (\ref{eq:ef_diskplanet}) and (\ref{eq:ef_planet}) show that  the forced planetesimal eccentricities in  massive  disks fall off more steeply with semimajor axis than in the massless disk case. Equally important is that massive disks suppress planetesimal eccentricities by more than an order of magnitude compared to massless disks, especially at $a \gg a_p$. 

\begin{figure}[t!]
\epsscale{1.17}
\plotone{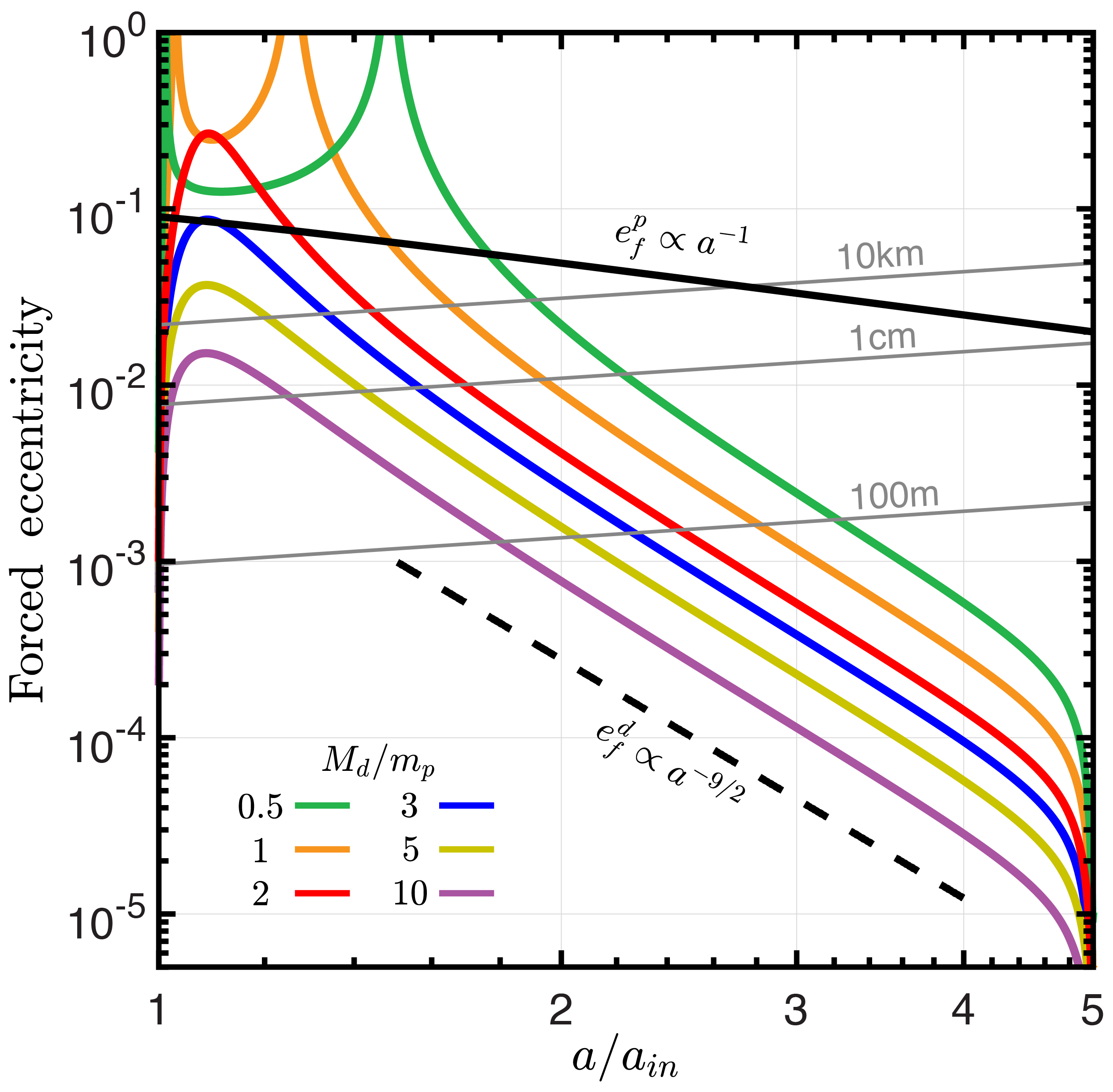}
\caption{Forced eccentricities of planetesimals $|e_{\textrm{f}}(a)|$ as a function of their semimajor axis (relative to $a\inn$). Calculations are done  for different values of $M_d/m_p$ using Equation (\ref{eq:e_forced}) for the same parameters as in Figure \ref{fig:precession-rates}. For reference, the black solid line represents the forced eccentricity in  a massless disk $e_{\rm{f}}^p$ (Equation (\ref{eq:ef_planet})), and the dashed line illustrates the asymptotic behavior of the eccentricities in massive disks $e_{\textrm{f}}^{d}$ (Equation (\ref{eq:ef_diskplanet})). All of these curves scale linearly with the planetary eccentricity $e_p$, here taken to be $0.10$. The thin gray lines represent the minimum eccentricities $e_{\textrm{stir}}$ for particles of radii 1cm, 100m and 10km to be stirred (Equation (\ref{eq:e_stir_full})). The disk is considered to be unstirred at distances where $|e_{\textrm{f}}(a)| \lesssim e_{\rm{stir}}(1 \textrm{cm})$. See the text (Sections \ref{sec:planetesimal_eccentriticies} and \ref{sec:implication_for_stirring_debris}) for details.}
\label{fig:eccentricities}
\end{figure}

This is illustrated in Figure \ref{fig:eccentricities}, where we plot the radial profiles of planetesimal forced eccentricities $e_{\textrm{f}}$ given by Equation (\ref{eq:e_forced}).  Calculations assume the same parameters as in Figure \ref{fig:precession-rates}. It is evident that for $M_d/m_p \gtrsim 1$, models ignoring the disk's gravitational effects -- so that, $e_{\rm{f}} = e_{\rm{f}}^p$ (Equation (\ref{eq:ef_planet}); solid black curve in Figure \ref{fig:eccentricities})  -- overestimate the planetesimal eccentricities by more than an order of magnitude throughout nearly the entire disk and $e_{\rm{f}} \rightarrow e_{\rm{f}}^{d}$ (Equation (\ref{eq:ef_diskplanet})). This is true even for disk masses $M_d/m_p \lesssim 1$; however, only in the outer regions of the disk, away from the location of the secular resonance where $e_{\rm{f}}$ diverges -- see Figure \ref{fig:eccentricities} and \citet{Paper1, Paper2}.

These findings have important consequences for stirring debris disks as we discuss further in Sections \ref{sec:implication_for_stirring_debris} and \ref{sec:discussion}. Before moving on, however, we note that a precessing planet inducing a smaller forced planetesimal eccentricity than a non-precessing one makes intuitive sense. In the extreme case of $\dot{\varpi}_p   \rightarrow \infty$, the eccentric planet's perturbations would reduce to that of a circular planet with an effective semimajor axis of $\approx Q_p$ \citep[][]{bat2011kbos, dawson_neptune}.

\subsection{Typical evolution/orbit-crossing timescales}

Equations (\ref{eq:e_t}) and (\ref{eq:w_t}) show that the timescale describing the periodic evolution of $e(t)$ and $\Delta\varpi(t)$ is $T_{\textrm{osc}}  = 2\pi/ |A_p +A_d-A_{d,p}|$. One can also easily show that $T_{\textrm{osc}}$ is related to the timescales at which debris particles, sharing the same semimajor axis, cross each other's orbits \citep[e.g.,][]{thebault2006, Mustill2009}. In the case where the dynamics is dominated by the disk gravity (i.e., $|A_d|, A_{d,p} \gtrsim A_p$), we 
find that $T_{\textrm{osc}} \rightarrow T_{\textrm{osc}}^{d}$ where: 
\begin{equation}
    T_{\textrm{osc}}^{d}
    \approx \frac{2\pi}{|A_d(a) - A_{d,p}|} 
    \approx      9 ~\textrm{Myr} ~\frac{3}{\phi_1^c} \bigg( \frac{1 M_N}{M_d} \bigg) 
    \frac{a_{\textrm{out}, 150}^{3}}{a_{p,30}^{3/2}} 
    M_{c,1}^{1/2}, 
    \label{eq:timescales}
\end{equation}
for our fiducial system (i.e., $p=3/2$, $\delta =5$). Note that the numerical estimate in Equation (\ref{eq:timescales}) ignores the contribution of $A_d(a)$ (recall that $|A_d(a)| \lesssim A_{d,p}$; see, e.g., Figure \ref{fig:precession-rates}). We find that this yields a satisfactory description, generally overestimating the exact value of $T_{\textrm{osc}}$ by a factor of $\lesssim 1.5$, particularly when $a\inn \lesssim a \lesssim a\out$, across all systems considered in Figures \ref{fig:precession-rates} and \ref{fig:eccentricities}. Note that  the approximation in Equation (\ref{eq:timescales}) implies that $ T_{\textrm{osc}}^d$ is independent of $a$; this holds roughly true even if $A_d(a)$ was not ignored. This is because the disk's finite extent renders $\psi_1 = \psi_1(a)$  \citep{sil15, SR19}, causing $|A_d(a)|$  to be slightly shallower than the $\propto a^{-1}$ profile given by the second line of Equation (\ref{eq:Ad}). In a massless disk, on the other hand, we find that \citep[see also, e.g.,][]{Mustill2009}:
\begin{equation}
   T_{\textrm{osc}} \rightarrow  T_{\textrm{osc}}^{p}
    = \frac{2\pi}{A_p(a)} 
    \approx      48 ~\textrm{Myr} ~ 
    ~ \bigg( \frac{1~ M_N}{m_p} \bigg)
    ~  \frac{a_{60}^{7/2}}{a_{p, 30}^{2}}  
    ~ M_{c,1}^{1/2}     .
    \label{eq:timescales_planet}
\end{equation}
Equations (\ref{eq:timescales}) and (\ref{eq:timescales_planet}) indicate that debris particles in relatively massive disks evolve, and consequently, cross each other's orbits more rapidly than in massless disks, particularly at distances $a \gg a_p$. This said, however, we note that the maximum  orbital eccentricities attained during the course of the evolution are significantly smaller in massive disks compared to massless disks (Section \ref{sec:planetesimal_eccentriticies}).

\section{Considerations of debris disk stirring}
\label{sec:implication_for_stirring_debris}

The findings outlined in Section \ref{sec:dynamics} have crucial implications for secular stirring of debris disks by planetary companions, which we now examine in detail.

\subsection{Relative collisional and fragmentation velocities}
\label{sec:relative_velocities}

We first  analyze the relative collisional velocities between planetesimals and characterize their fragmentation velocities.

Assuming that particles have randomized, uniformly distributed apsidal angles $\Delta\varpi$, the relative collisional velocity between particles can be written as   $v_{\rm rel}   = c  e   v_{\textrm K}$, where $v_{\textrm K} = n a$ is the local Keplerian orbital velocity, and $c$ is a constant of order unity  \citep{lissauer}. For values of $c \approx 1.4$ and $2$, with $e = | e_{\rm f}| $ (Equation (\ref{eq:e_forced})), $v_{\rm rel}$ could be understood as the mean and maximum relative velocities, respectively \citep[e.g.,][]{Mustill2009}.\footnote{For a Rayleigh distribution of eccentricities,  $c = \sqrt{5/4}$ for the mean relative velocities \citep[ignoring inclinations;][]{ida92}. That distribution, however, would arise due to the mutual gravitational scattering of planetesimals, which is different from the secular perturbations we consider. For more details, we refer the reader to \citet{Mustill2009}.} Given this, we can estimate the mean collisional velocities, $\langle v_{\textrm{rel}} \rangle$, in a massless disk, i.e., with $e = e_{\textrm{f}}^p$ (Equation (\ref{eq:ef_planet})), as follows: 
\begin{eqnarray}
\langle v_{\textrm{rel}}^p \rangle &\approx& 
156 ~ \textrm{m.s}^{-1} ~
M_{c,1}^{1/2} a_{100}^{-3/2} a_{p, 30} \frac{e_p}{0.10} .
\label{eq:vrel_planet}
\end{eqnarray}
In a massive disk ($M_d/m_p \gtrsim 1$), on the other hand, i.e., with $e = e_{\textrm{f}}^d$ (Equation (\ref{eq:ef_diskplanet})), we find that: 
\begin{eqnarray}
\langle v_{\textrm{rel}}^{d} \rangle &\approx& 4.8 ~  \textrm{m.s}^{-1} ~  M_{c,1}^{1/2} \frac{3}{\phi_1^c} \frac{m_p}{M_d} \frac{e_p}{0.10} \frac{a_{p,30}^{3/2} a_{\textrm{out}, 150}^3 }{a_{100}^5} ,
\label{eq:vrel_disk}
\end{eqnarray} 
assuming the fiducial disk parameters (i.e., $p=3/2$, $\delta =5$), and 
\begin{equation}
\langle v_{\textrm{rel}}^{d} \rangle / \langle v_{\textrm{rel}}^{p} \rangle   
= e_{\rm f}^{d}/e_{\rm f}^{p}  \approx  3.1 \times 10^{-2} ~  
\frac{3}{\phi_1^c} \frac{m_p}{M_d}  \frac{a_{p,30}^{1/2} a_{\textrm{out}, 150}^3 }{a_{100}^{7/2}}.
\label{eq:vrel_d_p_ratio}
\end{equation} 
Equations (\ref{eq:vrel_planet})--(\ref{eq:vrel_d_p_ratio}) show that the disk gravity lowers the relative collisional velocities of debris particles by more than an order of magnitude, especially at $a\gg a_p$. Such an effect was first pointed out by \citet{rafikov_ptype} in the context of planetesimal dynamics in circumbinary protoplanetary disks.

Generally speaking, collisions lead to fragmentation when the relative collisional velocities exceed a threshold value:
\begin{eqnarray}
 v_{\rm rel}
    &\gtrsim& v_{\textrm{frag}} .
    \label{eq:vrel_vfrag_equality}
\end{eqnarray} 
In principle, the fragmentation velocity $v_{\textrm{frag}}$ depends on several physical properties of the debris particles, including their radii $R_p$ \citep[e.g.,][]{housen90}. Here, and in line with previous studies, we define $v_{\textrm{frag}}$ as the minimum velocity at which collisions between  particles of \textit{basalt} material and of \textit{similar} radii no longer result in a net gain of mass \citep[e.g.,][and references therein]{wyattdent2002, krivov2005, krivov2018, Tyson2023}:
\begin{equation}
    v_{\textrm{frag}}(R_p) = 17.5 ~ {\textrm{m.s}}^{-1}  
    ~
    \bigg[ \bigg(\frac{R_p}{\textrm{m}}\bigg)^{-0.36} + \bigg( \frac{R_p}{{\textrm{km}}}\bigg)^{1.4}   \bigg]^{2/3} .
    \label{eq:vfrag}
\end{equation}
This ``V''-shaped function has a minimum of $\approx 6.56 \textrm{m.s}^{-1}$ at $R_p \approx 112$m, and $v_{\textrm{frag}}(1 \textrm{cm}) \approx v_{\textrm{frag}}(3.24 \textrm{km}) \approx 52.8 \textrm{m.s}^{-1}$. It is worth noting that, for the specific set of parameters in Equations (\ref{eq:vrel_planet}) and (\ref{eq:vrel_disk}), $\langle v_{\textrm{rel}}^{d} \rangle$ is lower than the minimum value of $v_{\textrm{frag}}(R_p)$, in contrast to $\langle v_{\textrm{rel}}^{p} \rangle$.

\subsection{Critical eccentricity for stirring}
\label{sec:critical_e}

We now estimate the minimum eccentricity required to stir  debris disks, and compare it to the forced planetesimal eccentricities with and without disk gravity.

To this end, we first translate the fragmentation condition (Equation (\ref{eq:vrel_vfrag_equality})) into a constraint on the minimum eccentricity, $e_{\textrm{stir}}$, required for stirring the orbits of debris particles:
\begin{equation}
        e_{\textrm{stir}}(a,R_p) \equiv  \frac{v_{\textrm{frag}}(R_p)}{c v_{\textrm{K}}(a)} 
        \approx 1.27 \times 10^{-3} ~ M_{c,1}^{-1/2} a_1^{1/2} . 
        \label{eq:e_stir_full}
\end{equation}
In Equation (\ref{eq:e_stir_full}),  we have used $c \approx 1.4$ (as before), and the numerical estimate assumes $R_p = 1$cm. The latter is chosen because observations probe sub-mm grains, thus larger particles  must be destructively colliding for the disk to be stirred and produce the observed dust \citep[see, e.g.,][]{Tyson2023}. Here, we note that the condition of Equation (\ref{eq:e_stir_full}) reduces to  the one presented in \citet{Tyson2023} upon setting $c = 2$.

As already discussed by \citet{Tyson2023}, debris particles with eccentricities below $e_{\textrm{stir}}(a, R_p)$ remain almost certainly unstirred, while those exceeding it might get stirred (depending on $\varpi(a)$, which is not accounted for here; see, e.g., \citet{Whitmire98}). For reference, the function $e_{\textrm{stir}}(a,R_p)$ given by Equation (\ref{eq:e_stir_full}) is overplotted in Figure \ref{fig:eccentricities} for three different values of particle radii: $R_p = 1 {\textrm{cm}}$,  $100 {\textrm{m}}$, and $10 {\textrm{km}}$. Looking at Figure \ref{fig:eccentricities}, it is evident that for $M_d/m_p \gtrsim 1$, the disk's gravity plays a protective role, preventing destructive collisions among debris particles of various sizes across a substantial portion of the disk's radial extent. Furthermore, based on the argument that debris disks will not produce observable dust if $e(a) \lesssim e_{\textrm{stir}}(1 {\textrm{cm}})$ \citep{Tyson2023}, Figure \ref{fig:eccentricities} indicates that massive disks would inhibit the expected secular stirring by eccentric planets  beyond $a/a\inn \sim 1 - 2 $, depending on the exact value of $M_d/m_p$.

\subsection{Critical semimajor axis demarcating the disk's (un)stirred region}
\label{sec:critical_a}

We now derive simple analytical formulae to characterize the distances beyond which the disk gravity hinders secular stirring and the production of observable dust.

The critical semimajor axis $a_{{\rm stir}}$ that demarcates the inner radius of the region where the disk gravity hinders destructive collisions/dust production can be estimated by equating $  e_{\rm f}^{d} $ and $e_{\rm stir}(1 {\textrm{cm}})$ (Equations (\ref{eq:ef_diskplanet}) and (\ref{eq:e_stir_full})). Doing so, we find:
\begin{eqnarray}
    \frac{a_{\rm stir}^{d}}{a\inn} & \approx &   1.43 
   ~ a_{\textrm{in}, 30}^{-1/10} ~
    \bigg[ 
    C \frac{p+1}{2-p} \frac{3}{\phi_1^c} \delta^3 \frac{e_p}{0.1} \frac{m_p}{M_d} \bigg(\frac{a_p}{a\inn}\bigg)^{3/2}
    M_{c,1}^{1/2}
    \bigg]^{1/5},
    \label{eq:a_stir}
    \\
    & \approx & 1.93 
    ~ 
    a_{\textrm{in}, 30}^{-1/10}
    ~  \bigg[ \frac{e_p}{0.1} \frac{m_p}{M_d} \bigg(\frac{a_p/a\inn}{0.8}\bigg)^{3/2}     \frac{3}{\phi_1^c}
    M_{c,1}^{1/2} \bigg]^{1/5} .
\nonumber
\end{eqnarray}
which, for $a_p \sim a\inn$ and $M_d/m_p \gtrsim 1$, is comparable to $a\inn$. Here, the second line assumes $p=3/2$ and $\delta = 5$, and it provides a good approximation to the true values of $a_{\rm stir}$ where $e_{\textrm{f}} = e_{\textrm{stir}}(1\textrm{cm})$ in Figure \ref{fig:eccentricities} (to within $\lesssim 5\%$). Note that according to Equation (\ref{eq:a_stir}) increasing (decreasing) $M_d/m_p$ at a fixed $a_p/a\inn$ shifts $a_{\rm{stir}}/a\inn$ to smaller (larger) values; see also Figure \ref{fig:eccentricities}. This makes intuitive sense as more massive disks drive faster planetary precession (Equation \ref{eq:Adp}). Equation (\ref{eq:a_stir}) is to compared with the massless disk case; setting $e_{\textrm{f}}^p = e_{\textrm{stir}}(1 {\textrm{cm}})$, we find that:
\begin{equation}
     \frac{a_{\rm stir}^p}{a\out} \approx  
    1.18 ~ 
    a_{\textrm{out}, 150}^{-1/3}
     \bigg[  \bigg( \frac{a_p/a\inn}{0.8} \bigg) \bigg(\frac{\delta}{5}\bigg)^{-1}  \frac{e_p}{0.1}   M_{c,1}^{1/2}  \bigg]^{2/3}, 
\end{equation}
which, for $a_p \sim a\inn$, is typically comparable to or  larger than $a\out \gtrsim a\inn$. Thus, planet-induced secular stirring of debris disks can be severely hindered by the disk's gravity, provided that $M_d \gtrsim m_p$ for planets near the disk's inner edge.

\subsection{Maximum radius of stirred debris particles}
\label{sec:max_Rp}

Finally, we compute the maximum sizes $R_{p, \textrm{max}}$ of debris particles that can be destroyed at a given semimajor axis, both in massless and massive disks.

To this end, we first note that for $R_p \gg 1$m,  Equation (\ref{eq:vfrag}) can be approximated as $v_{\textrm{frag}} \approx 17.5 ~\textrm{m.s}^{-1} (R_p/ \textrm{km})^{14/15}$. Combining this with the mean relative velocities in a massless disk (Equation (\ref{eq:vrel_planet})), we find that the size of the largest body that can be destroyed by the planet at a given $a$ is:
\begin{eqnarray}
   R_{p, \textrm{max}}^{p} &\approx& 18.7 ~ {\textrm{km}} ~  
   \bigg[ 
    \frac{e_p}{0.1} 
     \bigg(  \frac{a_p/a\inn}{0.8} \bigg)  
    \bigg(\frac{a/a\inn}{2}\bigg)^{-3/2} 
\times \nonumber \\ & & \qquad \quad \qquad \qquad \qquad
     a_{\textrm{in}, 30}^{-1/2}
    M_{c,1}^{1/2} 
   \bigg]^{15/14}.  
   \label{eq:Rmax_p}
\end{eqnarray}
Equation (\ref{eq:Rmax_p}) closely resembles the one derived by \citet[][Equation (25)]{Mustill2009}; however, there are slight numerical disparities stemming from variations in the definitions of $v_{\textrm{frag}}$ employed (see, e.g., their Equation (22)). In a massive disk, on the other hand, using Equation (\ref{eq:vrel_disk}) -- which, we remind, assumes $p = 3/2$ and $\delta = 5$ -- we find that the largest body that can be destroyed has a radius of 
\begin{eqnarray}
    R_{p, \textrm{max}}^{d} &\approx& 2.7 ~ {\textrm{km}} ~ 
    \bigg[ \frac{3}{\phi_1^c} \frac{m_p}{M_d} 
    \frac{e_p}{0.1} 
    \bigg(\frac{a_p/a\inn}{0.8}\bigg)^{3/2} 
    \times \nonumber \\ & & \qquad \qquad 
    \bigg(\frac{a/a\inn}{2}\bigg)^{-5} 
    a_{\textrm{in}, 30}^{-1/2}
    M_{c,1}^{1/2} 
   \bigg]^{15/14}.
    \label{eq:Rmax_d}
\end{eqnarray}
Looking at Equations (\ref{eq:Rmax_p}) and (\ref{eq:Rmax_d}), one can see that when collisions are destructive, the largest bodies contributing to the collisional cascade are smaller in massive disks compared to those in massless disks, i.e., $R_{p, \textrm{max}}^d / R_{p, \textrm{max}}^p \lesssim 1$. This is especially true for debris particles further away from the planet as $R_{p, \textrm{max}}^d / R_{p, \textrm{max}}^p \propto (a/a\inn)^{-7/2}$.  This behavior can also be seen by comparing the curves for forced planetesimal eccentricities, i.e., $e_{\textrm{f}}^d(a)$ and $e_{\textrm{f}}^p(a)$, with those for $e_{\textrm{stir}}(R_p)$ in Figure \ref{fig:eccentricities}, noting that  $e_{\textrm{stir}}(1 \textrm{cm}) \approx e_{\textrm{stir}}(3.24 \textrm{km})$; see Section \ref{sec:relative_velocities}.
These findings could have important consequences for  modeling of collisional cascades and particle size distributions in debris disks \citep[e.g.,][]{krivovwyatt20}.

\begin{table*}[ht!]
\begin{center}
\caption{Planetary systems with observed debris disks and inferred planets based on secular-stirring arguments.
\label{tab:table_1}}
\begin{tabular}{lccccccc}
\tableline
\tableline
System$^{(a)}$ & $M_c$ & $t_{\textrm{age}}$ & $[a\inn;  a\out ]^{(b)}$ & $m_p$ & $a_p$ & $a_{\textrm{stir}}^{(c)}$  & $m_p^{\textrm{min}}$
\\
 & $(M_{\odot})$ & (Myr) & (au) & $(M_J)$ & (au)   &  (au) & $(M_J)$
\\
\tableline
HD 32297  & 1.68 & 100 & $[91; 153] $  &  0.04 &  70     &   ... --  124.8  &  0.63  
\\
HD 121617 & 1.90 & 16  & $[50; 106] $  & 0.30   & 37      &   75.9 -- 100.7  &  1.00  
\\
HD 131835 & 1.81 & 16  & $[40; 127]$   & 1.20  &  30     &    91.5 --  122    &  3.24 
\\
HD 146181 & 1.28 & 16  & $[70;  90 ]$   & 0.04  &   53    &   ... -- 78.1    &  0.35  
\\
HD 192425 & 1.94 & 400 & $[100; 420 ]$ & 0.80  &   80   &  212.6 -- 332.1    &   10.65  
\\ 
HD 218396 & 1.59 & 42  & $[170; 210] $ & 0.05  & 120      &  ... -- ...     &     0.66   
\\
\tableline
\end{tabular}
\tablecomments{For each system, the values of $M_c$, $t_{\textrm{age}}$, and the disk's extent, i.e., $a\inn$ and $a\out$, are taken from \citet[][ignoring the reported observational uncertainties]{Pearce2022JWST}. Columns 5 and 6 present the planetary masses $m_p$ and semimajor axes $a_p$, as inferred by \citet[][where the authors assume $e_p = 0.30$]{Pearce2022JWST} based on secular stirring models of \textit{massless} disks \citep{Mustill2009}. Column 7 presents the minimum and maximum semimajor axis $a_{\textrm{stir}}$ up to which the inferred planets can stir the disks, considering disk masses of $M_d = 300 M_{\earth}$ and $30 M_{\earth}$, respectively. Column 8 presents the minimum planetary masses $m_p^{\textrm{min}}$ required to stir the  disks completely (specifically, up to $0.95 a\out$), assuming  $M_d = 100 M_{\earth}$ and the same values of $a_p$. See the text (Section \ref{sec:discussion}) for further details. 
(a). All systems, except HD 218396 (HR 8799), are not known to have planets. For HD 218396, as in \citet{Pearce2022JWST}, we ignore the presence of the known planets. 
(b). Each disk is resolved by ALMA at millimeter wavelengths except HD 192425, which is resolved by \textit{Herschel} at $0.1$mm. The lower spatial resolution of \textit{Herschel} means that the disk width of HD 192425 could be overestimated, affecting the planetary inferences  in both massless and massive debris disk models.
(c). An empty  cell, ``...", indicates that the entire disk remains unstirred.}
\end{center}
\end{table*}

\section{Discussion and Conclusions}
\label{sec:discussion}

Debris disks are often used as probes of exo-planetary systems, as their locations, characteristics and morphologies offer invaluable insights into the presence and properties of planets within their host systems \citep[][]{hughes2018review, wyatt19review}. In this spirit, a recent study  by \citet{Pearce2022JWST} analyzed a large sample of debris disk-bearing systems ($178$ in number). Through the application of inner-edge sculpting and secular planet-stirring arguments, the study concluded that the majority of debris disks require $\sim$Neptune- to Saturn-mass planets within their inner edges, typically at $\sim 10-100$ au, with only a few cases requiring $\sim$Jupiter-mass planets (see, e.g., Figure 14 in \citet{Pearce2022JWST}). While intriguing, the majority of the predicted planets with sub-Jupiter masses remain beyond the reach of current instruments, including JWST (\citet{Pearce2022JWST}; see also \citet{carter-jwst}).

As discussed in Section \ref{sec:intro}, planet inferences based on stirring models often use (analytical) models that consider only the gravity of the invoked planet(s), treating the disks as massless entities, as is also done in, e.g., \citet{Pearce2022JWST}. The ``traditional" planet-stirring mechanism considered in the literature for such purposes is the secular stirring by eccentric planets \citep{Mustill2009}. In this context, however, our results suggest that in systems with more massive debris disks than planets, the secular stirring expected from the planet is reduced, both in severity and radial range,  and may even be inhibited across the disk for $M_d/m_p \gg 1$ (Sections \ref{sec:dynamics} and \ref{sec:implication_for_stirring_debris}). Indeed, if we naively assume that the typical mass of a debris disk is $\sim 100 - 1000 M_{\earth}$ \citep{krivovwyatt20}, our results imply that the disk cannot be secularly stirred except if the planet is more massive than that, i.e., $\gtrsim 0.3 - 3 M_J$, thus surpassing the values predicted for the majority of planets in \citet{Pearce2022JWST}. If true, this implies that there could be many more systems within the sample examined by \citet{Pearce2022JWST} hosting planets with sufficient mass for potential detection.\footnote{For instance, JWST is expected to be sensitive to sub-Jupiter (sub-Saturn) mass planets beyond $\sim 30$ au ($\sim 50$ au), and as small as $0.1 M_J$ at $\gtrsim 100$ au \citep[see, e.g.,][]{carter-jwst}.} This should be taken into consideration when, e.g., planning planet observations.\footnote{A remark to this effect has been made by \citet{Pearce2022JWST}, where the authors discuss several caveats inherent to the results therein.}

To  further illustrate this point, Table \ref{tab:table_1} lists the values of $a_{\textrm{stir}}$ for six systems known to host debris disks assuming $M_d \gtrsim 0$, computed by  solving the  condition $|e_{\textrm{f}}(a)| = e_{\textrm{stir}}(1 \textrm{cm})$  numerically (Section \ref{sec:critical_a}). These systems are resolved at long wavelengths, which primarily trace large grains that are largely insensitive to radiation forces, thus serving as a proxy for the distribution of parent planetesimals. Calculations are done adopting the values of $m_p$ and $a_p$ that have been inferred by \citet[][]{Pearce2022JWST}  using secular planet-stirring models of \textit{massless} disks (\citet{Mustill2009}; see Columns 5 and 6 in Table \ref{tab:table_1}). These combinations of $m_p$ and $a_p$, together with $e_p = 0.30$ -- as stipulated by \citet{Pearce2022JWST} -- would stir the disks in Table \ref{tab:table_1} up to their observed outer edges if $M_d = 0$ (i.e., $|e_{\textrm{f}}(a)| = e_{\textrm{f}}^p(a) \gtrsim  e_{\textrm{stir}}(1 \textrm{cm})$ at $a = a\out$ in our notation).  It is natural then to ask how this situation would be altered if the disks are assumed to have $M_d \neq 0$. This is shown in Column 7 of Table \ref{tab:table_1}, where we report the minimum and maximum values of $a_{\textrm{stir}}$ corresponding to $M_d = 300 M_{\earth}$ and $30 M_{\earth}$, respectively.\footnote{Note that the systems in Table \ref{tab:table_1} are shown to require unfeasible disk masses of $\gtrsim$1000$M_{\earth}$  for self-stirring to be efficient \citep[][see their Table 1]{Pearce2022JWST}; values that are much larger than the maximum $M_d$ we consider.}  One can see that in all systems, the disk gravity shifts $a_{\textrm{stir}}$ inwards, i.e., closer to $a\inn$, and may even inhibit stirring across the entire disk.

Conversely, assuming a maximum disk mass of $M_d = 100 M_{\earth}$ (chosen arbitrarily), we can also compute the minimum planetary masses $m_p^{\textrm{min}}$ required to secularly stir the disks in Table \ref{tab:table_1}. We perform such a calculation such that the disks would be stirred up to\footnote{This is done to prevent the overestimation of $m_p^{\textrm{min}}$ in light of the divergence of $A_d$ at $a = a\out$, which causes $e_{\textrm{f}}(a\out) \rightarrow 0$; see Figures \ref{fig:precession-rates} and \ref{fig:eccentricities}.} $0.95 a\out$ (while keeping the values of $a_p$ unchanged; more on this below). The results are shown in the last column of Table \ref{tab:table_1}. Comparing $m_p^{\textrm{min}}$ and $m_p$, it is evident that the planetary masses need to be considerably larger (by up to a factor of $\sim 10$) than currently thought based on massless disk models. Before moving on, however, we would like to stress that our objective here is not to refine or predict the planetary parameters for these systems, but rather to provide a proof-of-concept illustrating the role of disk gravity. Partly for this reason, when calculating $m_p^{\textrm{min}}$, we utilized the same values of $a_p$ as those inferred by \citet{Pearce2022JWST}. Nevertheless, we note that  $m_p^{\textrm{min}}$ can be roughly interpreted as the minimum planetary mass with a maximum semimajor axis of $a_p$. This can be understood as follows. According to Equation (\ref{eq:a_stir}) -- which, recall, assumes $A_{d,p} \gtrsim |A_d|, A_p$ -- if the planet were to have a smaller semimajor axis, while all other factors remain constant, maintaining the same value of $a_{\rm stir}^{d}$  would require $m_p (a_p/a\inn)^{3/2}$ to remain constant (ignoring for simplicity the dependence of $\phi_1^c$ on $a_p/a\inn$; Section \ref{sec:basicphysics}).

Taken at face value,  the results of preceding sections also imply that if $M_d \gtrsim m_p$, then systems with low, e.g., sub-Saturn mass planets would exhibit an anti-correlation (correlation) with the presence of (relatively narrow) debris disks. Conversely, the presence or future detection of planets in debris-bearing systems can be translated into constraints on the total mass of debris disks for them to be stirred (see Equations (\ref{eq:Mdisk_condition}) and (\ref{eq:a_stir})). Indeed, assuming that the observed outer edge of a given disk corresponds to the semimajor axis of the outermost stirred particle, i.e., $a_{\textrm{stir}} = a\out$, Equation (\ref{eq:a_stir}) would then yield a \textit{maximum} debris disk mass. Namely, assuming $\delta \gg 1$ and $0<p<2$ (i.e., most of the disk mass is in the outer regions; Equation (\ref{eq:Sigma_d})), we find that $M_d \propto e_p m_p (a_p/a\inn)^{3/2} / \delta^{3+p}$. The physical significance of such constraints in addressing the ``debris disk mass problem" \citep{krivovwyatt20} described in Section \ref{sec:intro} would naturally depend on the  planetary parameters.

To exemplify this, we consider the $\beta$ Pic planetary system. According to ALMA observations \citep{matra-beta-pic-hotcold}, its debris disk extends from $a\inn \approx 25$ au to $a\out \approx 150$ au (i.e., $\delta \approx 6$). Assuming $p=3/2$ (Equation (\ref{eq:Sigma_d})), we solve the condition $|e_{\textrm{f}}(a)| = e_{\textrm{stir}}(1 \textrm{cm})$  numerically (Section \ref{sec:critical_a}), finding a maximum disk mass of $M_d \approx 7 M_{\earth}$ and $\approx 14 M_{\earth}$ upon stipulating $a_{\textrm{stir}}/a\out = 0.95$ and $0.90$, respectively. In these simple calculations, we only considered the outer of the two planets in the system, i.e., $\beta$ Pic b (with $m_p = 9.3 M_J$, $a_p = 10.26$au, $e_p = 0.12$, and $M_c = 1.83 M_{\odot}$; \citet{Brandt}), ignoring the presence of $\beta$ Pic c at $\approx 2.7$ au \citep{lagrange-betapic-c}. Thus, the reported values of $M_d$ serve as a first-order minimum estimate. Remarkably, however, these values of $M_d$ are in agreement with the minimum disk mass estimated  by \citet[][]{krivovwyatt20} -- namely,  $\approx 5.5 M_{\earth}$ -- based on collisional models assuming the largest planetesimal size is $0.3$ km.

Last but not least, it is worth emphasizing that the analytical estimates provided in this paper primarily pertain to radially broad disks with $\delta = a\out/a\inn = 5$. However, it is essential to recognize that the fundamental physics extends to wider and even narrower disks, albeit with some simple numerical adjustments. Notably, the disks in Table 1 exhibit $\delta$ ranging between $\approx 1.2$ to $\approx 4.2$, showcasing the versatility of the considered physics across a spectrum of disk widths. In this regard, the first line of Equation (\ref{eq:a_stir}) provides a handy tool to gauge the outer boundaries within which secular planet-stirring could be effective in massive debris disks. However, we note that caution must be exercised especially when considering relatively narrow disks ($\delta \approx 1$). This is because mean motion resonances (MMRs), which we have ignored, could play an important dynamical role (see also the next paragraph), questioning the validity of secular approximation. For instance, for a Jupiter-mass planet,  the nominal location of the 2:1 MMR -- namely, $a_{2:1} = 2^{2/3} a_p$ (i.e., neglecting potential offsets due to the disk-induced precession; \citet{Murray2022}) -- would be within the disk at $a_{2:1}/a\inn \approx 1.07$ if $e_p = 0.10$, but  $a_{2:1} \lesssim a\inn$  if $e_p = 0.30$ (assuming $M_c = 1 M_{\odot}$ and using Equation (\ref{eq:rHQ}) to    relate $a_p$ and $a\inn$). 

Finally, several factors may influence our conclusions. 
First, we focused solely on the axisymmetric contribution of the disk gravity, neglecting its non-axisymmetric component. This enabled us to analytically model the disk's secular effects. While the non-axisymmetric gravitational potential ($\propto B_d$) induced by the debris disk's eccentricity could theoretically impact the stirring process, we expect minimal influence in the considered regime ($M_d/m_p \gtrsim 1$ and $a_p/a\inn \lesssim 1$). This is because, by definition, $B_d \propto e_d$ \citep{sil15, ST19}, justifying our assumptions when $e_d \sim e_{\textrm{f}}^{d} \ll e_p$.
The exchange of angular momentum between disk particles (due to $B_d \neq 0$) could, however, cause some departure of $e(a)$  from the $\propto a^{-9/2}$ profile of Equation (\ref{eq:ef_diskplanet}). 
These issues should be addressed in the future, and they can be handled in a computationally effective manner through, e.g., the use of existing orbit-averaged, secular codes designed for simulating the evolution self-gravitating disks. Examples include (i) the \texttt{Gauss} code of \citet{JTgauss}, which is valid to all orders of eccentricities and inclinations, and (ii) the linear \texttt{N-RING} code of \citet{Paper2} -- valid to second order in eccentricities -- based on \citet{hahn2003}. 
Second, we considered only secular (coplanar) interactions, ignoring the possibility of multiple (possibly inclined) planets, mean motion resonances and scattering with the planet(s), as well as self-stirring. These processes may indeed contribute to the stirring process  \citep[as studied in the case of the Kuiper belt, e.g.,][]{malhotra95, wardhahn1998, gladman2006, dawson_neptune, sousa2019}, and their collective impact in a self-gravitating disk warrants further investigation.\footnote{Notably, the gravitational coupling between planets and self-gravitating disks at resonances facilitates the exchange of energy and/or angular momentum \citep{goldreich79, goldreich80}. For instance, mean motion or Lindblad resonances may increase the planetary eccentricity  $\dot{e}_p \gtrsim 0$ and induce migration toward the star $\dot{a}_p \lesssim 0$ \citep{goldreich2003}, while secular resonances may damp $e_p$ without affecting $a_p$ \citep{tre98, wardhahn1998, wardhahnprotostars, hahn2008, Paper2}. Such variations in the planetary orbit would then affect the disk stirring levels. Exploring these effects warrants a comprehensive investigation in the future (J. Touma \& A. Sefilian, in prep.) and lies beyond the scope of this work.} 
Third, we employed a simple collisional and stirring criterion (i.e., Equation (\ref{eq:e_stir_full}) with $R_p = 1$cm); more detailed models \citep[such as those in, e.g.,][]{lohne2008, silsbee2021, Tyson2023} may reveal some differences when compared to our predictions. 
Fourth and finally, we considered gas-free disks; some debris disks harbor significant amounts of gas \citep{hughes2018review}, which may affect the dynamics considered here. We defer detailed investigation of these issues to future works. 
Nevertheless, the findings of the present study clearly indicate that disk gravity can significantly impact long-term planet-debris disk interactions, emphasizing the need to address a substantial groundwork in this regard.

To summarize, however, such details are unlikely to undermine the illustrated proof of concept here: 
relatively massive debris disks  -- through their axisymmetric gravitational potential --
may hinder the expected secular stirring process due to  planets on eccentric orbits \citep[e.g.,][]{Mustill2009}.
At best, this finding holds important implications for planetary inferences derived from stirring analyses of debris disks. 
At worst, our findings could necessitate adjustments to the masses and orbital parameters of planets invoked to replicate observed levels of stirring, particularly when relying on secular stirring models of massless disks \citep[e.g.,][]{Mustill2009, Pearce2022JWST}. Conversely, if planets are known and/or detected in debris-bearing systems, the findings of this  study could be utilized to indirectly measure the total masses of debris disks.

In closing, we remark that the present study is inspired by and hence a variation on the works of  \citet{rafikov_ptype, rafikov_stype} concerning planetesimal dynamics in massive, circumstellar/binary protoplanetary disks.
Indeed, similar to how massive, nearly axisymmetric protoplanetary disks may propel planetesimals past fragmentation barriers and facilitate growth in stellar binaries, relatively massive debris disks may hinder secular stirring by eccentric planets and impede dust production through collisions.

\section*{Acknowledgments}
\noindent  A.A.S. is supported by the Alexander von Humboldt Foundation through a Humboldt Research Fellowship for postdoctoral researchers. Back-of-the-envelope calculations for this study were carried out during the Ph.D. research of A.A.S. at DAMTP, University of Cambridge, supported by funding from the Gates Cambridge Trust. The paper was completed during a research visit to Pontificia Universidad Católica de Chile in January 2024, under the generous hosting of Cristobal Petrovich. A.A.S. is grateful to Jihad Touma and Cristobal Petrovich for their engaging discussions and constructive comments on earlier drafts, which significantly improved the overall quality and clarity of the work.
A.A.S. also thanks Roman Rafikov and Tim Pearce for their comments on the final draft, as well as the anonymous referee for their positive report and constructive comments on the manuscript.

\section*{Data availability}
\noindent The data supporting the findings of this study are available within the article.


\bibliography{Sefilian2024Stirring}{}
\bibliographystyle{aasjournal}


\end{document}